\documentclass[a4paper,10pt]{article}
\usepackage[utf8]{inputenc}
\usepackage[top=0.95in, bottom=0.75in, left=1.28in, right=0.8in]{geometry}
\usepackage{mathtools,hyperref}
\usepackage{tikz}
\usepackage{graphics}
\usepackage{setspace}
\title{\large \textbf {Quadrature Squeezing with Normally Ordered Noise Operators}}
\author{\large Merid Tufa* and Fesseha Kassahun\\Department of Physics, Addis Ababa University, P. O. Box 1176, Addis Ababa, Ethiopia}
\author{Merid Tufa \footnote{Email address: meridtufa@gmail.com} and 
Fesseha Kassahun\footnote{Email address: fessehakassahun@gmail.com}\\
                Department of Physics,
                Addis Ababa University\\
                P. O. Box 1176, Addis Ababa, Ethiopia}
\begin{document}
\maketitle
\begin{abstract}
We have considered the interaction of subharmonic light modes with a three-level atom in a closed cavity coupled to a vacuum reservoir. We carry out analysis by normally ordering the vacuum reservoir noise operators. It so happens that there is perfect quadrature squeezing under certain conditions. 
\end{abstract}
\providecommand{\keywords}[1]{\textbf{{keywords:}} #1}
\section{Introduction}
There has been a considerable interest in the analysis of the quantum properties of the squeezed light generated by
various quantum optical systems such as subharmonic generators and three-level atoms[1-10]. In squeezed light the noise in one quadrature is below the vacuum-state level at the expense of enhanced fluctuations in the conjugate quadrature, with the product of the uncertainties in the two quadratures satisfying the uncertainty relation [10]. In addition to exhibiting  nonclassical features, squeezed light has potential applications in low-noise optical communications, precision measurements, and weak signal detections [2, 3].\\
\indent
It has been found that three-level atoms in a closed cavity coupled to vacuum reservoir and pumped by electron bombardment or coherent light generate squeezed light [9, 10]. It has been also shown theoretically [10-13] and subsequently confirmed experimentally [14-16] that subharmonic generation produces squeezed light, with a maximum quadrature squeezing of $50\%$ below the vacuum-state level [10].\\
\indent
In addition, some authors have studied the statistical and squeezing properties of the light generated by three-level atoms interacting with subharmonic light modes, using the usual commutation relation [17-20].However, it appears to be difficult to believe the results obtained in this manner to be correct in light of the discussion given in Ref. [21].\\
\indent
In this paper, we wish to analyze the quadrature squeezing of the two-mode cavity light produced by the interaction of subharmonic light modes (emerging from a nonlinear crystal pumped by coherent light) with a three-level atom by putting the vacuum reservoir noise operators in normal order. We consider the case in which the nonlinear crystal and the three-level atom are in a closed cavity coupled to a vacuum reservoir via a single port mirror. We carry out our analysis applying the master equation for the cavity modes and atomic operators. The large-time approximation is used to decouple the equations of evolution of cavity modes and atomic operators. Finally, employing the steady-state solutions of the resulting equations of evolution for the expectation values of the cavity modes and atomic operators, we calculate the quadrature squeezing by normally ordering the  vacuum reservoir noise operators.

\section{Operator Dynamics}
\noindent 
We consider the case in which a nonlinear crystal driven by coherent light and a three-level atom are available in a closed cavity coupled to a vacuum reservoir. Moreover, we consider the case in which the top and bottom levels of the three-level atom are not coupled by the coherent light emerging from the nonlinear crystal. This is physically realized by covering the right-side of the nonlinear crystal by a screen which can absorb the coherent light. We denote the top, intermediate, and bottom levels of the atom by $ | a\rangle, |b\rangle$ and $ |c\rangle$, respectively.\\ 
\indent
The process of subharmonic generation taking place inside the nonlinear crystal can be described by the  Hamiltonian
\begin{equation}\label{1}
 \hat{H}' = i\lambda(\hat{c}^{\dagger}\hat{a}\hat{b} - \hat{c}\hat{a}^{\dagger}\hat{b}^{\dagger}),
\end{equation}
where the operators $\hat{a}$ and $\hat{b}$ represent the subharmonic light modes, $\lambda$ is the coupling constant between the coherent light and light mode a or b and $\hat{c}$ is the annihilation operator for the coherent light. In order to have a manageable mathematical analysis, we replace the operator $\hat{c}$ by $\gamma$ which is taken to be real, positive, and constant, we can write the Hamiltonian as
\begin{equation}\label{2}
 \hat{H}' = i\varepsilon(\hat{a}\hat{b} - \hat{a}^{\dagger}\hat{b}^{\dagger}),
\end{equation}
where $ \varepsilon = \lambda\gamma$.
\begin{figure*}
 \centering
 \begin{center}
 \includegraphics[width=8cm, height = 5cm]{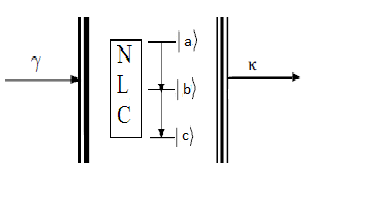}
  \caption{A three-level atom with a nonlinear crystal (NLC).}
 \end{center}
\end{figure*}
 In addition, the interaction of the subharmonic light modes with the three-level atom can be described at resonance by the Hamiltonian
\begin{equation}\label{3}
 \hat{H}'' = ig(\hat{\sigma}^{\dagger}_a\hat{a} - \hat{a}^{\dagger}\hat{\sigma}_a + \hat{\sigma}^{\dagger}_b\hat{b} - \hat{b}^{\dagger}\hat{\sigma}_b),
\end{equation}
where $g$ is the coupling constant between the atom and light mode a or b and $\hat{\sigma}_a$ and $\hat{\sigma}_b$ are lowering atomic operators defined by
\begin{equation}\label{4}
 \hat{\sigma}_a = |b\rangle\langle a|,
\end{equation}
\begin{equation}\label{5}
 \hat{\sigma}_b = |c\rangle\langle b|.
\end{equation}
\noindent
Thus the interactions involving the cavity light modes are described by
\begin{equation}\label{6}
 \hat{H_c} = i\varepsilon(\hat{a}\hat{b} - \hat{a}^{\dagger}\hat{b}^{\dagger}) + ig(\hat{\sigma}^{\dagger}_a\hat{a} - \hat{a}^{\dagger}\hat{\sigma}_a + \hat{\sigma}^{\dagger}_b\hat{b} - \hat{b}^{\dagger}\hat{\sigma}_b)
\end{equation}
and the interaction involving the three-atom is given by
\begin{equation}\label{7}
\hat{H}_a = ig(\hat{\sigma}^{\dagger}_a\hat{a} - \hat{a}^{\dagger}\hat{\sigma}_a + \hat{\sigma}^{\dagger}_b\hat{b} - \hat{b}^{\dagger}\hat{\sigma}_b).
\end{equation}
\indent
Now we seek to obtain the equations of evolution for the cavity mode operators employing the master equation 
\begin{equation}\label{8}
 \frac{d}{dt}\hat{\rho}(t) = - i\big[\hat{H}_c, \hat{\rho}\big]  + \frac{\kappa}{2}\bigg(2\hat{a}\rho\hat{a}^{\dagger} - \hat{a}^{\dagger}\hat{a}\rho - \rho\hat{a}^{\dagger}\hat{a}\bigg) + \frac{\kappa}{2}\bigg(2\hat{b}\rho\hat{b}^{\dagger} - \hat{b}^{\dagger}\hat{b}\rho - \rho\hat{b}^{\dagger}\hat{b}\bigg).
\end{equation}
On account of Eq. \eqref{6}, we have
\begin{eqnarray}\label{9}
 \frac{d}{dt}\hat{\rho}(t) &=& - i\big[i\varepsilon(\hat{a}\hat{b} - \hat{a}^{\dagger}\hat{b}^{\dagger}) + ig(\hat{\sigma}^{\dagger}_a\hat{a} - \hat{a}^{\dagger}\hat{\sigma}_a + \hat{\sigma}^{\dagger}_b\hat{b} - \hat{b}^{\dagger}\hat{\sigma}_b), \hat{\rho}\big]  + \frac{\kappa}{2}\bigg(2\hat{a}\rho\hat{a}^{\dagger} - \hat{a}^{\dagger}\hat{a}\rho\nonumber\\
 &-& \rho\hat{a}^{\dagger}\hat{a}\bigg) + \frac{\kappa}{2}\bigg(2\hat{b}\rho\hat{b}^{\dagger} - \hat{b}^{\dagger}\hat{b}\rho - \rho\hat{b}^{\dagger}\hat{b}\bigg).
\end{eqnarray}
We recall that the equation of evolution for an operator $\hat{A}$ is expressible as
\begin{equation}\label{10}
 \frac{d}{dt}\langle\hat{A}\rangle = Tr\bigg(\frac{d}{dt}\hat{\rho}(t)\hat{A}\bigg).
\end{equation}
Then employing this relation along with Eq. \eqref{9}, we readily obtain
\begin{equation}\label{11}
 \frac{d}{dt}\langle\hat{a}(t)\rangle = -\frac{\kappa}{2}\langle\hat{a}(t)\rangle - \varepsilon\langle\hat{b}^{\dagger}(t)\rangle - g\langle\hat{\sigma}_a(t)\rangle,
\end{equation}
\begin{equation}\label{12}
 \frac{d}{dt}\langle\hat{b}(t)\rangle = -\frac{\kappa}{2}\langle\hat{b}(t)\rangle - \varepsilon\langle\hat{a}^{\dagger}(t)\rangle - g\langle\hat{\sigma}_b(t)\rangle,
\end{equation}
\begin{equation}\label{13}
 \frac{d}{dt}\langle\hat{a}^{\dagger}(t)\hat{a}(t)\rangle = -\kappa\langle\hat{a}^{\dagger}(t)\hat{a}(t)\rangle - \varepsilon\big\langle\hat{b}(t)\hat{a}(t) +\hat{a}^{\dagger}(t)\hat{b}^{\dagger}(t)\big\rangle - g\big\langle\hat{\sigma}^{\dagger}_a(t)\hat{a}(t) + \hat{a}^{\dagger}(t)\hat{\sigma}_a(t)\big\rangle,
\end{equation}
\begin{equation}\label{14}
 \frac{d}{dt}\langle\hat{a}(t)\hat{a}^{\dagger}(t)\rangle = -\kappa\langle\hat{a}(t)\hat{a}^{\dagger}(t)\rangle - \varepsilon\big\langle\hat{a}(t)\hat{b}(t) +\hat{b}^{\dagger}(t)\hat{a}^{\dagger}(t)\big\rangle - g\big\langle\hat{a}(t)\hat{\sigma}^{\dagger}_a(t) + \hat{\sigma}_a(t)\hat{a}^{\dagger}(t)\big\rangle + \kappa,
\end{equation}
\begin{equation}\label{15}
 \frac{d}{dt}\langle\hat{b}^{\dagger}(t)\hat{b}(t)\rangle = -\kappa\langle\hat{b}^{\dagger}(t)\hat{b}(t)\rangle - \varepsilon\big\langle\hat{a}(t)\hat{b}(t) +\hat{b}^{\dagger}(t)\hat{a}^{\dagger}(t)\big\rangle - g\big\langle\hat{\sigma}^{\dagger}_b(t)\hat{b}(t) + \hat{b}^{\dagger}(t)\hat{\sigma}_b(t)\big\rangle,
\end{equation}
\begin{equation}\label{16}
 \frac{d}{dt}\langle\hat{b}(t)\hat{b}^{\dagger}(t)\rangle = -\kappa\langle\hat{b}(t)\hat{b}^{\dagger}(t)\rangle - \varepsilon\big\langle\hat{b}(t)\hat{a}(t) +\hat{a}^{\dagger}(t)\hat{b}^{\dagger}(t)\big\rangle - g\big\langle\hat{b}(t)\hat{\sigma}^{\dagger}_b(t) + \hat{\sigma}_b(t)\hat{b}^{\dagger}(t)\big\rangle + \kappa,
\end{equation}
\begin{equation}\label{17}
\frac{d}{dt}\langle\hat{a}(t)\hat{b}(t)\rangle = -\kappa\langle\hat{a}(t)\hat{b}(t)\rangle - \varepsilon\big\langle\hat{a}^{\dagger}(t)\hat{a}(t) +\hat{b}^{\dagger}(t)\hat{b}(t)\big\rangle - g\big\langle\hat{\sigma}_a(t)\hat{b}(t) + \hat{a}(t)\hat{\sigma}_b(t)\big\rangle - \varepsilon,
\end{equation}
\begin{equation}\label{18}
\frac{d}{dt}\langle\hat{b}(t)\hat{a}(t)\rangle = -\kappa\langle\hat{b}(t)\hat{a}(t)\rangle - \varepsilon\big\langle\hat{a}^{\dagger}(t)\hat{a}(t) +\hat{b}^{\dagger}(t)\hat{b}(t)\big\rangle - g\big\langle\hat{b}(t)\hat{\sigma}_a(t) + \hat{\sigma}_b(t)\hat{a}(t)\big\rangle - \varepsilon,
\end{equation}
\begin{equation}\label{19}
\frac{d}{dt}\langle\hat{a}^{2}(t)\rangle = -\kappa\langle\hat{a}^2(t)\rangle - \varepsilon\big\langle\hat{a}(t)\hat{b}^{\dagger}(t) +\hat{b}^{\dagger}(t)\hat{a}(t)\big\rangle - g\big\langle\hat{a}(t)\hat{\sigma}_a(t) + \hat{\sigma}_a(t)\hat{a}(t)\big\rangle,
\end{equation}
\begin{equation}\label{20}
\frac{d}{dt}\langle\hat{b}^{2}(t)\rangle = -\kappa\langle\hat{b}^2(t)\rangle - \varepsilon\big\langle\hat{b}(t)\hat{a}^{\dagger}(t) +\hat{a}^{\dagger}(t)\hat{b}(t)\big\rangle - g\big\langle\hat{b}(t)\hat{\sigma}_b(t) + \hat{\sigma}_b(t)\hat{b}(t)\big\rangle,
\end{equation}
\begin{equation}\label{21}
 \frac{d}{dt}\langle\hat{a}^{\dagger}(t)\hat{b}(t)\rangle = -\kappa\langle\hat{a}^{\dagger}(t)\hat{b}(t)\rangle - \varepsilon\big\langle\hat{a}^{\dagger 2}(t) +\hat{b}^{2}(t)\big\rangle - g\big\langle\hat{\sigma}^{\dagger}_a(t)\hat{b}(t) + \hat{a}^{\dagger}(t)\hat{\sigma}_b(t)\big\rangle,
\end{equation}
and
\begin{equation}\label{22}
 \frac{d}{dt}\langle\hat{b}(t)\hat{a}^{\dagger}(t)\rangle = -\kappa\langle\hat{b}(t)\hat{a}^{\dagger}(t)\rangle - \varepsilon\big\langle\hat{a}^{\dagger 2}(t) +\hat{b}^{2}(t)\big\rangle - g\big\langle\hat{b}(t)\hat{\sigma}^{\dagger}_a(t) + \hat{\sigma}_b(t)\hat{a}^{\dagger}(t)\big\rangle.
\end{equation}
\indent
We see that Eqs. \eqref{13}-\eqref{22} are nonlinear differential equations and hence it is not possible to find the exact time-dependent solutions of these equations. We intend to overcome this problem by applying the large-time approximation scheme [18,19]To this end, we rewrite Eqs. \eqref{11} and \eqref{12} as 
\begin{equation}\label{23}
 \frac{d}{dt}\hat{a}(t) = -\frac{\kappa}{2}\hat{a}(t) -\varepsilon\hat{b}^{\dagger}(t)
 - g\hat{\sigma}_a(t) + \hat{F}_a(t),
\end{equation}
\begin{equation}\label{24}
\frac{d}{dt}\hat{b}(t) = -\frac{\kappa}{2}\hat{b}(t) -\varepsilon\hat{a}^{\dagger}(t)
 - g\hat{\sigma}_b(t) + \hat{F}_b(t),
\end{equation}
where $\hat{F}_a(t)$ and $\hat{F}_b(t)$ are noise operators with vanishing mean and associated with the cavity mode operators $\hat{a}(t)$ and $\hat{b}(t)$, respectively. Then applying the large-time approximation scheme to these equations, we get the following approximately valid relations
\begin{equation}\label{25}
 \hat{a}(t) = -\frac{2\varepsilon}{\kappa}\hat{b}^{\dagger}(t) - \frac{2g}{\kappa}\hat{\sigma}_a(t) + \frac{2}{\kappa}\hat{F}_a(t)
\end{equation}
and
\begin{equation}\label{26}
 \hat{b}(t) = -\frac{2\varepsilon}{\kappa}\hat{a}^{\dagger}(t) - \frac{2g}{\kappa}\hat{\sigma}_b(t) + \frac{2}{\kappa}\hat{F}_b(t).
\end{equation}
It then follows that
\begin{equation}\label{27}
 \hat{a}(t) = \frac{4\varepsilon\kappa g}{\kappa^2 - 4\varepsilon^2}\bigg(\frac{\hat{\sigma}^{\dagger}_b(t)}{\kappa} - \frac{\hat{\sigma}_a(t)}{2\varepsilon} + \frac{\hat{F}_a(t)}{2\varepsilon g} - \frac{\hat{F}^{\dagger}_b(t)}{\kappa g}\bigg)
\end{equation}
and
\begin{equation}\label{28}
 \hat{b}(t) = \frac{4\varepsilon\kappa g}{\kappa^2 - 4\varepsilon^2}\bigg(\frac{\hat{\sigma}^{\dagger}_a(t)}{\kappa} - \frac{\hat{\sigma}_b(t)}{2\varepsilon} + \frac{\hat{F}_b(t)}{2\varepsilon g} - \frac{\hat{F}^{\dagger}_a(t)}{\kappa g}\bigg).
\end{equation}
Upon substituting Eqs. \eqref{27} and \eqref{28} together with their adjoint into Eqs. \eqref{13}-\eqref{22}, we readily obtain
\begin{eqnarray}\label{29}
 \frac{d}{dt}\langle\hat{a}^{\dagger}(t)\hat{a}(t)\rangle &=& -\kappa\langle\hat{a}^{\dagger}(t)\hat{a}(t)\rangle - \varepsilon\big\langle\hat{b}(t)\hat{a}(t) +\hat{a}^{\dagger}(t)\hat{b}^{\dagger}(t)\big\rangle - \frac{4\varepsilon\kappa g^2}{\kappa^2 - 4\varepsilon^2}\bigg[\frac{\langle\hat{\sigma}_c(t) + \hat{\sigma}^{\dagger}_c(t)\rangle}{\kappa}\nonumber\\ 
 &&- \frac{\langle\hat{\eta}_a(t)\rangle}{\varepsilon} + \bigg\langle\hat{\sigma}^{\dagger}_a(t) \bigg(\frac{\hat{F}_a(t)}{2\varepsilon g} - \frac{\hat{F}^{\dagger}_b(t)}{\kappa g}\bigg) + \bigg(\frac{\hat{F}^{\dagger}_a(t)}{2\varepsilon g} - \frac{\hat{F}_b(t)}{\kappa g}\bigg)\hat{\sigma}_a(t)\bigg\rangle\bigg],
\end{eqnarray}
\begin{eqnarray}\label{30}
 \frac{d}{dt}\langle\hat{a}(t)\hat{a}^{\dagger}(t)\rangle &=& -\kappa\langle\hat{a}(t)\hat{a}^{\dagger}(t)\rangle - \varepsilon\big\langle\hat{a}(t)\hat{b}(t) +\hat{b}^{\dagger}(t)\hat{a}^{\dagger}(t)\big\rangle - \frac{4\varepsilon\kappa g^2}{\kappa^2 - 4\varepsilon^2}\bigg[-\frac{\langle\hat{\eta}_b(t)\rangle}{\varepsilon}\nonumber\\
 &&+ \bigg\langle\bigg(\frac{\hat{F}_a(t)}{2\varepsilon g} - \frac{\hat{F}^{\dagger}_b(t)}{\kappa g}\bigg)\hat{\sigma}^{\dagger}_a(t)  + \hat{\sigma}_a(t)\bigg(\frac{\hat{F}^{\dagger}_a(t)}{2\varepsilon g} - \frac{\hat{F}_b(t)}{\kappa g}\bigg)\bigg\rangle\bigg] + \kappa,
\end{eqnarray}
\begin{eqnarray}\label{31}
 \frac{d}{dt}\langle\hat{b}^{\dagger}(t)\hat{b}(t)\rangle &=& -\kappa\langle\hat{b}^{\dagger}(t)\hat{b}(t)\rangle - \varepsilon\big\langle\hat{a}(t)\hat{b}(t) +\hat{b}^{\dagger}(t)\hat{a}^{\dagger}(t)\big\rangle - \frac{4\varepsilon\kappa g^2}{\kappa^2 - 4\varepsilon^2}\bigg[-\frac{\langle\hat{\eta}_b(t)\rangle}{\varepsilon}\nonumber\\
&&+ \bigg\langle\hat{\sigma}^{\dagger}_b(t) \bigg(\frac{\hat{F}_b(t)}{2\varepsilon g} - \frac{\hat{F}^{\dagger}_a(t)}{\kappa g}\bigg) + \bigg(\frac{\hat{F}^{\dagger}_b(t)}{2\varepsilon g} - \frac{\hat{F}_a(t)}{\kappa g}\bigg)\hat{\sigma}_b(t)\bigg\rangle\bigg],
\end{eqnarray}
\begin{eqnarray}\label{32}
 \frac{d}{dt}\langle\hat{b}(t)\hat{b}^{\dagger}(t)\rangle &=& -\kappa\langle\hat{b}(t)\hat{b}^{\dagger}(t)\rangle - \varepsilon\big\langle\hat{b}(t)\hat{a}(t) +\hat{a}^{\dagger}(t)\hat{b}^{\dagger}(t)\big\rangle - \frac{4\varepsilon\kappa g^2}{\kappa^2 - 4\varepsilon^2}\bigg[\frac{\langle\hat{\sigma}_c(t) + \hat{\sigma}^{\dagger}_c(t)\rangle}{\kappa}\nonumber\\
 &&- \frac{\langle\hat{\eta}_c(t)\rangle}{\varepsilon} + \bigg\langle\bigg(\frac{\hat{F}_b(t)}{2\varepsilon g} - \frac{\hat{F}^{\dagger}_a(t)}{\kappa g}\bigg)\hat{\sigma}^{\dagger}_b(t)  + \hat{\sigma}_b(t)\bigg(\frac{\hat{F}^{\dagger}_b(t)}{2\varepsilon g} - \frac{\hat{F}_a(t)}{\kappa g}\bigg)\bigg\rangle\bigg]\nonumber\\&& + \kappa,
\end{eqnarray}
\begin{eqnarray}\label{33}
\frac{d}{dt}\langle\hat{a}(t)\hat{b}(t)\rangle &=& -\kappa\langle\hat{a}(t)\hat{b}(t)\rangle - \varepsilon\big\langle\hat{a}^{\dagger}(t)\hat{a}(t) +\hat{b}^{\dagger}(t)\hat{b}(t)\big\rangle - \frac{4\varepsilon\kappa g^2}{\kappa^2 - 4\varepsilon^2}\bigg[\frac{2\langle\hat{\eta}_b(t)\rangle}{\kappa}\nonumber\\
&&+ \bigg\langle\hat{\sigma}_a(t) \bigg(\frac{\hat{F}_b(t)}{2\varepsilon g} - \frac{\hat{F}^{\dagger}_a(t)}{\kappa g}\bigg) + \bigg(\frac{\hat{F}_a(t)}{2\varepsilon g} - \frac{\hat{F}^{\dagger}_b(t)}{\kappa g}\bigg)\hat{\sigma}_b(t)\bigg\rangle\bigg] -\varepsilon,
\end{eqnarray}
\begin{eqnarray}\label{34}
\frac{d}{dt}\langle\hat{b}(t)\hat{a}(t)\rangle &=& -\kappa\langle\hat{b}(t)\hat{a}(t)\rangle - \varepsilon\big\langle\hat{a}^{\dagger}(t)\hat{a}(t) +\hat{b}^{\dagger}(t)\hat{b}(t)\big\rangle - \frac{4\varepsilon\kappa g^2}{\kappa^2 - 4\varepsilon^2}\bigg[\frac{\langle\hat{\eta}_a(t) + \hat{\eta}_c(t)\rangle}{\kappa}\nonumber\\
 &&- \frac{\langle\hat{\sigma}_c(t)\rangle}{\varepsilon} + \bigg\langle\bigg(\frac{\hat{F}_b(t)}{2\varepsilon g} - \frac{\hat{F}^{\dagger}_a(t)}{\kappa g}\bigg)\hat{\sigma}_a(t)  + \hat{\sigma}_b(t)\bigg(\frac{\hat{F}_a(t)}{2\varepsilon g} - \frac{\hat{F}^{\dagger}_b(t)}{\kappa g}\bigg)\bigg\rangle\bigg]\nonumber\\&&  -\varepsilon,
\end{eqnarray}
\begin{eqnarray}\label{35}
\frac{d}{dt}\langle\hat{a}^{2}(t)\rangle&=&-\kappa\langle\hat{a}^2(t)\rangle - \varepsilon\big\langle\hat{a}(t)\hat{b}^{\dagger}(t) +\hat{b}^{\dagger}(t)\hat{a}(t)\big\rangle - \frac{4\varepsilon\kappa g^2}{\kappa^2 - 4\varepsilon^2}\bigg[\bigg\langle\bigg(\frac{\hat{F}_a(t)}{2\varepsilon g} - \frac{\hat{F}^{\dagger}_b(t)}{\kappa g}\bigg)\hat{\sigma}_a(t)\nonumber\\&&+\hat{\sigma}_a(t)\bigg(\frac{\hat{F}_a(t)}{2\varepsilon g} - \frac{\hat{F}^{\dagger}_b(t)}{\kappa g}\bigg)\bigg\rangle\bigg],
\end{eqnarray}
\begin{eqnarray}\label{36}
\frac{d}{dt}\langle\hat{b}^{2}(t)\rangle &=& -\kappa\langle\hat{b}^2(t)\rangle - \varepsilon\big\langle\hat{b}(t)\hat{a}^{\dagger}(t) +\hat{a}^{\dagger}(t)\hat{b}(t)\big\rangle - \frac{4\varepsilon\kappa g^2}{\kappa^2 - 4\varepsilon^2}\bigg[\bigg\langle \bigg(\frac{\hat{F}_b(t)}{2\varepsilon g} - \frac{\hat{F}^{\dagger}_a(t)}{\kappa g}\bigg)\hat{\sigma}_b(t)\nonumber\\
&&+\hat{\sigma}_b(t)\bigg(\frac{\hat{F}_b(t)}{2\varepsilon g} - \frac{\hat{F}^{\dagger}_a(t)}{\kappa g}\bigg)\bigg\rangle\bigg],
\end{eqnarray}
\begin{eqnarray}\label{37}
 \frac{d}{dt}\langle\hat{a}^{\dagger}(t)\hat{b}(t)\rangle &=& -\kappa\langle\hat{a}^{\dagger}(t)\hat{b}(t)\rangle - \varepsilon\big\langle\hat{a}^{\dagger 2}(t) +\hat{b}^{2}(t)\big\rangle - \frac{4\varepsilon\kappa g^2}{\kappa^2 - 4\varepsilon^2}\bigg[\bigg\langle \hat{\sigma}^{\dagger}_a(t) \bigg(\frac{\hat{F}_b(t)}{2\varepsilon g} - \frac{\hat{F}^{\dagger}_a(t)}{\kappa g}\bigg)\nonumber\\
 &&+ \bigg(\frac{\hat{F}^{\dagger}_a(t)}{2\varepsilon g} - \frac{\hat{F}_b(t)}{\kappa g}\bigg)\hat{\sigma}_b(t)\bigg\rangle\bigg],
\end{eqnarray}
\begin{eqnarray}\label{38}
 \frac{d}{dt}\langle\hat{b}(t)\hat{a}^{\dagger}(t)\rangle &=& -\kappa\langle\hat{b}(t)\hat{a}^{\dagger}(t)\rangle - \varepsilon\big\langle\hat{a}^{\dagger 2}(t) +\hat{b}^{2}(t)\big\rangle - \frac{4\varepsilon\kappa g^2}{\kappa^2 - 4\varepsilon^2}\bigg[\bigg\langle \bigg(\frac{\hat{F}_b(t)}{2\varepsilon g} - \frac{\hat{F}^{\dagger}_a(t)}{\kappa g}\bigg)\hat{\sigma}^{\dagger}_a(t) \nonumber\\
 &&+\hat{\sigma}_b(t) \bigg(\frac{\hat{F}^{\dagger}_a(t)}{2\varepsilon g} - \frac{\hat{F}_b(t)}{\kappa g}\bigg)\bigg\rangle\bigg],
\end{eqnarray}
where
\begin{equation}\label{39}
 \hat{\sigma}_c = |c\rangle\langle a|,
\end{equation}
\begin{equation}\label{40}
 \hat{\eta}_a = |a\rangle\langle a|,
\end{equation}
\begin{equation}\label{41}
 \hat{\eta}_b = |b\rangle\langle b|,
\end{equation}
and
\begin{equation}\label{42}
 \hat{\eta}_c = |c\rangle\langle c|.
\end{equation}
Since an atomic operator and a noise operator associated with a cavity mode are not correlated, one can write
\begin{equation}\label{43}
 \langle\hat{F}(t)\hat{\sigma}(t)\rangle = \langle\hat{F}(t)\rangle\langle\hat{\sigma}(t)\rangle = 0
\end{equation}
and hence Eqs. \eqref{29}-\eqref{38} take the form
\begin{eqnarray}\label{44}
 \frac{d}{dt}\langle\hat{a}^{\dagger}(t)\hat{a}(t)\rangle &=& -\kappa\langle\hat{a}^{\dagger}(t)\hat{a}(t)\rangle - \varepsilon\big\langle\hat{b}(t)\hat{a}(t) +\hat{a}^{\dagger}(t)\hat{b}^{\dagger}(t)\big\rangle - \frac{4\varepsilon\kappa g^2}{\kappa^2 - 4\varepsilon^2}\bigg[\frac{\langle\hat{\sigma}_c(t) + \hat{\sigma}^{\dagger}_c(t)\rangle}{\kappa} \nonumber\\&&- \frac{\langle\hat{\eta}_a(t)\rangle}{\varepsilon}\bigg],
\end{eqnarray}
\begin{eqnarray}\label{45}
 \frac{d}{dt}\langle\hat{a}(t)\hat{a}^{\dagger}(t)\rangle &=& -\kappa\langle\hat{a}(t)\hat{a}^{\dagger}(t)\rangle - \varepsilon\big\langle\hat{a}(t)\hat{b}(t) +\hat{b}^{\dagger}(t)\hat{a}^{\dagger}(t)\big\rangle + \frac{4\varepsilon\kappa g^2}{\kappa^2 - 4\varepsilon^2}\frac{\langle\hat{\eta}_b(t)\rangle}{\varepsilon}\nonumber\\
 &&+ \kappa,
\end{eqnarray}
\begin{eqnarray}\label{46}
 \frac{d}{dt}\langle\hat{b}^{\dagger}(t)\hat{b}(t)\rangle = -\kappa\langle\hat{b}^{\dagger}(t)\hat{b}(t)\rangle - \varepsilon\big\langle\hat{a}(t)\hat{b}(t) +\hat{b}^{\dagger}(t)\hat{a}^{\dagger}(t)\big\rangle + \frac{4\varepsilon\kappa g^2}{\kappa^2 - 4\varepsilon^2}\frac{\hat{\langle\eta}_b(t)\rangle}{\varepsilon},
\end{eqnarray}
\begin{eqnarray}\label{47}
 \frac{d}{dt}\langle\hat{b}(t)\hat{b}^{\dagger}(t)\rangle &=& -\kappa\langle\hat{b}(t)\hat{b}^{\dagger}(t)\rangle - \varepsilon\big\langle\hat{b}(t)\hat{a}(t) +\hat{a}^{\dagger}(t)\hat{b}^{\dagger}(t)\big\rangle 
 - \frac{4\varepsilon\kappa g^2}{\kappa^2 - 4\varepsilon^2}\bigg[\frac{\langle\hat{\sigma}_c(t) + \hat{\sigma}^{\dagger}_c(t)\rangle}{\kappa}\nonumber\\&& - \frac{\langle\hat{\eta}_c(t)\rangle}{\varepsilon}\bigg] + \kappa,
\end{eqnarray}
\begin{eqnarray}\label{48}
\frac{d}{dt}\langle\hat{a}(t)\hat{b}(t)\rangle &=& -\kappa\langle\hat{a}(t)\hat{b}(t)\rangle - \varepsilon\big\langle\hat{a}^{\dagger}(t)\hat{a}(t) +\hat{b}^{\dagger}(t)\hat{b}(t)\big\rangle - \frac{4\varepsilon\kappa g^2}{\kappa^2 - 4\varepsilon^2}\frac{2\langle\hat{\eta}_b(t)\rangle}{\kappa} \nonumber\\&&- \varepsilon,
\end{eqnarray}
\begin{eqnarray}\label{49}
\frac{d}{dt}\langle\hat{b}(t)\hat{a}(t)\rangle &=& -\kappa\langle\hat{b}(t)\hat{a}(t)\rangle - \varepsilon\big\langle\hat{a}^{\dagger}(t)\hat{a}(t) +\hat{b}^{\dagger}(t)\hat{b}(t)\big\rangle - \frac{4\varepsilon\kappa g^2}{\kappa^2 - 4\varepsilon^2}\bigg[\frac{\langle\hat{\eta}_a(t) + \hat{\eta}_c(t)\rangle}{\kappa}
 \nonumber\\&&- \frac{\langle\hat{\sigma}_c(t)\rangle}{\varepsilon}\bigg] -\varepsilon,
\end{eqnarray}
\begin{eqnarray}\label{50}
\frac{d}{dt}\langle\hat{a}^{2}(t)\rangle = -\kappa\langle\hat{a}^2(t)\rangle - \varepsilon\big\langle\hat{a}(t)\hat{b}^{\dagger}(t) +\hat{b}^{\dagger}(t)\hat{a}(t)\big\rangle,
\end{eqnarray}
\begin{eqnarray}\label{51}
\frac{d}{dt}\langle\hat{b}^{2}(t)\rangle = -\kappa\langle\hat{b}^2(t)\rangle - \varepsilon\big\langle\hat{b}(t)\hat{a}^{\dagger}(t) +\hat{a}^{\dagger}(t)\hat{b}(t)\big\rangle,
\end{eqnarray}
\begin{eqnarray}\label{52}
 \frac{d}{dt}\langle\hat{a}^{\dagger}(t)\hat{b}(t)\rangle = -\kappa\langle\hat{a}^{\dagger}(t)\hat{b}(t)\rangle - \varepsilon\big\langle\hat{a}^{\dagger 2}(t) +\hat{b}^{2}(t)\big\rangle.
\end{eqnarray}
and
\begin{eqnarray}\label{53}
 \frac{d}{dt}\langle\hat{b}(t)\hat{a}^{\dagger}(t)\rangle = -\kappa\langle\hat{b}(t)\hat{a}^{\dagger}(t)\rangle - \varepsilon\big\langle\hat{a}^{\dagger 2}(t) +\hat{b}^{2}(t)\big\rangle.
\end{eqnarray}
The steady-state solutions of Eqs. \eqref{44}-\eqref{53} are found to be
\begin{eqnarray}\label{54}
\langle\hat{a}^{\dagger}\hat{a}\rangle = -\frac{\varepsilon}{\kappa}\big\langle\hat{b}\hat{a} + \hat{a}^{\dagger}\hat{b}^{\dagger}\big\rangle - \frac{4g^2}{\kappa^2 - 4\varepsilon^2}\bigg[\frac{\varepsilon\langle\hat{\sigma}_c + \hat{\sigma}^{\dagger}_c\rangle}{\kappa} - \langle\hat{\eta}_a\rangle\bigg],
\end{eqnarray}
\begin{eqnarray}\label{55}
\langle\hat{a}\hat{a}^{\dagger}\rangle = -\frac{\varepsilon}{\kappa}\big\langle\hat{a}\hat{b} +\hat{b}^{\dagger}\hat{a}^{\dagger}\big\rangle + \frac{4g^2}{\kappa^2 - 4\varepsilon^2}\langle\hat{\eta}_b\rangle +1,
\end{eqnarray}
\begin{eqnarray}\label{56}
 \langle\hat{b}^{\dagger}\hat{b}\rangle = -\frac{\varepsilon}{\kappa}\big\langle\hat{a}\hat{b} +\hat{b}^{\dagger}\hat{a}^{\dagger}\big\rangle + \frac{4g^2}{\kappa^2 - 4\varepsilon^2}\langle\hat{\eta}_b\rangle,
\end{eqnarray}
\begin{eqnarray}\label{57}
\langle\hat{b}\hat{b}^{\dagger}\rangle = -\frac{\varepsilon}{\kappa}\big\langle\hat{b}\hat{a} +\hat{a}^{\dagger}\hat{b}^{\dagger}\big\rangle - \frac{4g^2}{\kappa^2 - 4\varepsilon^2}\bigg[\frac{\varepsilon\langle\hat{\sigma}_c + \hat{\sigma}^{\dagger}_c\rangle}{\kappa} - \langle\hat{\eta}_c\rangle\bigg] + 1,
\end{eqnarray}
\begin{eqnarray}\label{58}
\langle\hat{a}\hat{b}\rangle = -\frac{\varepsilon}{\kappa}\big\langle\hat{a}^{\dagger}\hat{a} + \hat{b}^{\dagger}\hat{b}\big\rangle - \frac{4g^2}{\kappa^2 - 4\varepsilon^2}\frac{2\varepsilon\langle\hat{\eta}_b\rangle}{\kappa} - \frac{\varepsilon}{\kappa},
\end{eqnarray}
\begin{eqnarray}\label{59}
\langle\hat{b}\hat{a}\rangle = -\frac{\varepsilon}{\kappa}\big\langle\hat{a}^{\dagger}\hat{a} +\hat{b}^{\dagger}\hat{b}\big\rangle - \frac{4g^2}{\kappa^2 - 4\varepsilon^2}\bigg[\frac{\varepsilon\langle\hat{\eta}_a + \hat{\eta}_c\rangle}{\kappa}  -\langle\hat{\sigma}_c\rangle\bigg] -\frac{\varepsilon}{\kappa},
\end{eqnarray}
\begin{eqnarray}\label{60}
\langle\hat{a}^2\rangle = -\frac{\varepsilon}{\kappa}\big\langle\hat{a}\hat{b}^{\dagger} +\hat{b}^{\dagger}\hat{a}\big\rangle,
\end{eqnarray}
\begin{eqnarray}\label{61}
\langle\hat{b}^2\rangle = -\frac{\varepsilon}{\kappa}\big\langle\hat{b}\hat{a}^{\dagger} +\hat{a}^{\dagger}\hat{b}\big\rangle,
\end{eqnarray}
\begin{eqnarray}\label{62}
\langle\hat{a}^{\dagger}\hat{b}\rangle = -\frac{\varepsilon}{\kappa}\big\langle\hat{a}^{\dagger 2} +\hat{b}^{2}\big\rangle,
\end{eqnarray}
and 
\begin{eqnarray}\label{63}
\langle\hat{b}\hat{a}^{\dagger}\rangle = -\frac{\varepsilon}{\kappa}\big\langle\hat{a}^{\dagger 2} +\hat{b}^{2}\big\rangle.
\end{eqnarray}
Taking the complex conjugate of Eq. \eqref{60}, we have
\begin{eqnarray}\label{64}
\langle\hat{a}^{\dagger2}\rangle = -\frac{\varepsilon}{\kappa}\big\langle\hat{b}\hat{a}^{\dagger} +\hat{a}^{\dagger}\hat{b}\big\rangle.
\end{eqnarray}
In view of Eqs. \eqref{61} and \eqref{64}, Eq. \eqref{62} takes the form
\begin{eqnarray}\label{65}
\langle\hat{a}^{\dagger}\hat{b}\rangle = \frac{2\varepsilon^2}{\kappa^2}\big\langle\hat{b}\hat{a}^{\dagger} +\hat{a}^{\dagger}\hat{b}\big\rangle
\end{eqnarray} 
and also on account of Eqs. \eqref{62} and \eqref{63}, we have
\begin{eqnarray}\label{66}
\langle\hat{a}^{\dagger}\hat{b}\rangle = \frac{4\varepsilon^2}{\kappa^2}\big\langle\hat{a}^{\dagger}\hat{b}\big\rangle.
\end{eqnarray}  
This shows that
\begin{eqnarray}\label{67}
\langle\hat{a}^{\dagger}\hat{b}\rangle = \big\langle\hat{b}\hat{a}^{\dagger}\rangle = 0.
\end{eqnarray} 
In view of these results and their complex conjugates, Eqs. \eqref{60}, \eqref{61} and \eqref{64} turn out to be
\begin{eqnarray}\label{68}
\langle\hat{a}^2\rangle = \langle\hat{b}^2\rangle = \langle\hat{a}^{\dagger2}\rangle = \langle\hat{b}^{\dagger2}\rangle = 0.
\end{eqnarray}
Using Eqs.\eqref{58} and \eqref{59} along with their complex conjugates in Eqs. \eqref{54} and \eqref{56}, we easily find
\begin{eqnarray}\label{69}
 \langle\hat{a}^{\dagger}\hat{a}\rangle &=& \frac{2\varepsilon^2}{\kappa^2 - 2\varepsilon^2} + \frac{2\varepsilon^2}{\kappa^2 - 2\varepsilon^2}\langle\hat{b}^{\dagger}\hat{b}\rangle + \frac{4g^2}{(\kappa^2 - 2\varepsilon^2)(\kappa^2 - 4\varepsilon^2)} \bigg(2\varepsilon^2\langle\hat{\eta}_c\rangle + \langle\hat{\eta}_a\rangle(2\varepsilon^2 + \kappa^2)\nonumber\\&& - 2\varepsilon\kappa\big\langle\hat{\sigma}_c + \hat{\sigma}_c^{\dagger}\big\rangle\bigg)
\end{eqnarray}
and 
\begin{eqnarray}\label{70}
 \langle\hat{b}^{\dagger}\hat{b}\rangle &=& \frac{2\varepsilon^2}{\kappa^2 - 2\varepsilon^2} + \frac{2\varepsilon^2}{\kappa^2 - 2\varepsilon^2}\langle\hat{a}^{\dagger}\hat{a}\rangle + \frac{4g^2}{(\kappa^2 - 2\varepsilon^2)(\kappa^2 - 4\varepsilon^2)}\bigg(4\varepsilon^2 + \kappa^2\bigg)\langle\hat{\eta}_b\rangle.
\end{eqnarray}
Using Eqs. \eqref{69} and \eqref{70}, we readily obtain
\begin{eqnarray}\label{71}
 \langle\hat{a}^{\dagger}\hat{a}\rangle &=& \frac{2\varepsilon^2}{\kappa^2 - 4\varepsilon^2} + \frac{4g^2}{\kappa^2(\kappa^2 - 4\varepsilon^2)^2}\bigg((\kappa^4 - 4\varepsilon^4 )\langle\hat{\eta}_a\rangle + 2\varepsilon^2(4\varepsilon^2 + \kappa^2)\langle\hat{\eta}_b\rangle\nonumber\\&& +2\varepsilon^2(\kappa^2 - 2\varepsilon^2)\langle\hat{\eta}_c\rangle - 2\varepsilon\kappa(\kappa^2 - 2\varepsilon^2)\big\langle\hat{\sigma}_c + \hat{\sigma}^{\dagger}_c\big\rangle\bigg)
\end{eqnarray}
and 
\begin{eqnarray}\label{72}
 \langle\hat{b}^{\dagger}\hat{b}\rangle &=&\frac{2\varepsilon^2}{\kappa^2 - 4\varepsilon^2} + \frac{4g^2}{\kappa^2(\kappa^2 - 4\varepsilon^2)^2}\bigg(2\varepsilon^2(\kappa^2 + 2\varepsilon^2)\langle\hat{\eta}_a\rangle + (\kappa^2 + 4\varepsilon^2)(\kappa^2 - 2\varepsilon^2)\langle\hat{\eta}_b\rangle\nonumber\\&&+ 4\varepsilon^4\langle\hat{\eta}_c\rangle - 4\varepsilon^3\kappa\big\langle\hat{\sigma}_c + \hat{\sigma}^{\dagger}_c\big\rangle\bigg).
\end{eqnarray}
Furthermore, taking into account Eqs. \eqref{58} and \eqref{59} along with their complex conjugates, Eqs. \eqref{55} and \eqref{57} can be put in the form
\begin{eqnarray}\label{73}
\langle\hat{a}\hat{a}^{\dagger}\rangle &=& 1+ \frac{2\varepsilon^2}{\kappa^2 - 4\varepsilon^2} + \frac{4g^2}{\kappa^2(\kappa^2 - 4\varepsilon^2)^2}\bigg(2\varepsilon^2(\kappa^2 + 2\varepsilon^2)\langle\hat{\eta}_a\rangle + (\kappa^2 + 4\varepsilon^2)(\kappa^2 - 2\varepsilon^2)\langle\hat{\eta}_b\rangle\nonumber\\&&+ 4\varepsilon^4\langle\hat{\eta}_c\rangle - 4\varepsilon^3\kappa\big\langle\hat{\sigma}_c + \hat{\sigma}^{\dagger}_c\big\rangle\bigg)
\end{eqnarray}
and
\begin{eqnarray}\label{74}
\langle\hat{b}\hat{b}^{\dagger}\rangle &=& 1 +  \frac{2\varepsilon^2}{\kappa^2 - 4\varepsilon^2} + \frac{4g^2}{\kappa^2(\kappa^2 - 4\varepsilon^2)^2}\bigg(4\varepsilon^2(\kappa^2 - \varepsilon^2)\langle\hat{\eta}_a\rangle + 2\varepsilon^2(4\varepsilon^2 + \kappa^2)\langle\hat{\eta}_b\rangle\nonumber\\&&+ (\kappa^2 - 2\varepsilon^2(\kappa^2 + 2\varepsilon^2))\langle\hat{\eta}_c\rangle - 2\varepsilon\kappa(\kappa^2 - 2\varepsilon^2)\big\langle\hat{\sigma}_c + \hat{\sigma}^{\dagger}_c\big\rangle\bigg).
\end{eqnarray}
\indent
The three-level atom inside the closed cavity doesn't interact with the vacuum reservoir outside the cavity. Therefore, the equation of evolution of the density operator for this atom has the form [22]
\begin{equation}\label{75}
  \frac{d}{dt}\hat{\rho}(t) = -i\big[\hat{H}_a, \hat{\rho}\big].
\end{equation}
The equation of evolution for the expectation value of an atomic operator $\hat{\sigma}(t)$ can be written as
\begin{equation}\label{76}
  \frac{d}{dt}\langle \hat{\sigma}(t)\rangle = Tr\bigg(\frac{d}{dt}\hat{\rho}(t)\hat{\sigma}\bigg),
\end{equation}
so that in view of Eq. \eqref{75}, there follows
\begin{equation}\label{77}
  \frac{d}{dt}\langle \hat{\sigma}(t)\rangle = -iTr\bigg(\big[\hat{H}_a, \hat{\rho}\big]\hat{\sigma}\bigg)
\end{equation}
Now one can establish that
\begin{equation}\label{78}
  \frac{d}{dt}\langle \hat{\sigma}(t)\rangle = -i\langle\big[\hat{\sigma}, \hat{H}_a\big]\rangle.
\end{equation}
Applying this relation along with Eq. \eqref{7}, we readily get
\begin{equation}\label{79}
 \frac{d}{dt}\langle \hat{\sigma}_a(t)\rangle = g\bigg\langle \big(\hat{\eta}_b(t) - \hat{\eta}_a(t)\big)\hat{a}(t) + \hat{b}^{\dagger}(t)\hat{\sigma}_c(t)\bigg\rangle,
\end{equation}
\begin{equation}\label{80}
 \frac{d}{dt}\langle \hat{\sigma}_b(t)\rangle = g\bigg\langle -\hat{a}^{\dagger}(t)\hat{\sigma}_c(t) + \big(\hat{\eta}_c(t) - \hat{\eta}_b(t)\big)\hat{b}(t)\bigg\rangle,
\end{equation}
\begin{equation}\label{81}
 \frac{d}{dt}\langle \hat{\sigma}_c(t)\rangle = g\bigg\langle \hat{\sigma}_b(t)\hat{a}(t) - \hat{\sigma}_a(t)\hat{b}(t)\bigg\rangle,
\end{equation}
\begin{equation}\label{82}
 \frac{d}{dt}\langle \hat{\eta}_a(t)\rangle = g\bigg\langle \hat{\sigma}^{\dagger}_a(t)\hat{a}(t) + \hat{a}^{\dagger}(t)\hat{\sigma}_a(t)\bigg\rangle,
\end{equation}
\begin{equation}\label{83}
 \frac{d}{dt}\langle \hat{\eta}_b(t)\rangle = g\bigg\langle \hat{\sigma}^{\dagger}_b(t)\hat{b}(t) + \hat{b}^{\dagger}(t)\hat{\sigma}_b(t) - \big(\hat{\sigma}^{\dagger}_a(t)\hat{a}(t) + \hat{a}^{\dagger}(t)\hat{\sigma}_a(t)\big)\bigg\rangle,
\end{equation}
\begin{equation}\label{84}
 \frac{d}{dt}\langle \hat{\eta}_c(t)\rangle = -g\bigg\langle \hat{\sigma}^{\dagger}_b(t)\hat{b}(t) + \hat{b}^{\dagger}(t)\hat{\sigma}_b(t)\bigg\rangle,
\end{equation}
In views of Eqs. \eqref{27}, \eqref{28}, and \eqref{43} together with their adjoint, Eqs. \eqref{79}-\eqref{84} take the form
\begin{equation}\label{85}
 \frac{d}{dt}\langle\hat{\sigma}_a(t)\rangle = -\frac{\kappa^2\varepsilon\gamma_c}{\kappa^2 - 4\varepsilon^2}\bigg\langle \frac{\hat{\sigma}_a(t)}{2\varepsilon} - \frac{\hat{\sigma}^{\dagger}_b(t)}{\kappa}\bigg\rangle,
\end{equation}
\begin{equation}\label{86}
 \frac{d}{dt}\langle \hat{\sigma}_b(t)\rangle = -\frac{\kappa^2\gamma_c}{\kappa^2 - 4\varepsilon^2}\langle\hat{\sigma}_b(t)\rangle, 
\end{equation}
\begin{equation}\label{87}
 \frac{d}{dt}\langle \hat{\sigma}_c(t)\rangle = \frac{\kappa^2\varepsilon\gamma_c}{4\varepsilon^2 - \kappa^2}\bigg\langle\frac{\hat{\sigma}_c(t)}{2\varepsilon} + \frac{1}{\kappa}\big(\hat{\eta}_b(t) - \hat{\eta}_c(t)\big)\bigg\rangle,
\end{equation}
\begin{equation}\label{88}
 \frac{d}{dt}\langle \hat{\eta}_a(t)\rangle = \frac{\kappa^2\varepsilon\gamma_c}{4\varepsilon^2 - \kappa^2}\bigg\langle-\frac{1}{\kappa}\big(\hat{\sigma}^{\dagger}_c(t) + \hat{\sigma}_c(t)\big) + \frac{\hat{\eta}_a}{\varepsilon}\bigg\rangle,
\end{equation}
\begin{equation}\label{89}
 \frac{d}{dt}\langle \hat{\eta}_b(t)\rangle = \frac{\kappa^2\varepsilon\gamma_c}{4\varepsilon^2 - \kappa^2}\bigg\langle\frac{1}{\kappa}\big(\hat{\sigma}^{\dagger}_c(t) + \hat{\sigma}_c(t)\big) + \frac{1}{\varepsilon}\big(\hat{\eta}_b(t) - \hat{\eta}_a(t)\big)\bigg\rangle,
\end{equation}
\begin{equation}\label{90}
 \frac{d}{dt}\langle \hat{\eta}_c(t)\rangle = -\frac{\kappa^2\varepsilon\gamma_c}{4\varepsilon^2 - \kappa^2}\bigg\langle \frac{\hat{\eta}_b(t)}{\epsilon}\bigg\rangle,
\end{equation}
where $\gamma_c = \frac{4g^2}{\kappa} $ is the stimulated emission decay constant.\\
\indent
We find the steady-state solutions of Eqs. \eqref{87}-\eqref{89} to be of the form
\begin{equation}\label{91}
 \langle\hat{\sigma}_c\rangle = \frac{2\varepsilon}{\kappa}\bigg(\langle\hat{\eta}_c \rangle - \langle\hat{\eta}_b\rangle\bigg),
\end{equation}
\begin{equation}\label{92}
 \langle\hat{\sigma}^{\dagger}_c\rangle + \langle\hat{\sigma}_c\rangle = \frac{\kappa}{\varepsilon}\langle\hat{\eta}_a\rangle,
\end{equation}
\begin{equation}\label{93}
\langle\hat{\eta}_a\rangle - \langle\hat{\eta}_b\rangle = \frac{\varepsilon}{\kappa} \bigg(\langle\hat{\sigma}^{\dagger}_c\rangle + \langle\hat{\sigma}_c\rangle\bigg).
\end{equation}
Now from Eqs. \eqref{92} and \eqref{93}, one can easily get
\begin{equation}\label{94}
 \langle\hat{\eta}_b\rangle = 0.
\end{equation}
Upon substituting Eq. \eqref{91} into Eq. \eqref{92} and taking into account Eq.\eqref{94}, we readily obtain
\begin{equation}\label{95}
 \langle\hat{\eta}_c\rangle = \frac{\kappa^2}{4\varepsilon^2}\langle\hat{\eta}_a\rangle,
\end{equation}
The completeness relation for the three-level atom is given by
\begin{equation}\label{96}
\langle\hat{\eta}_a\rangle +  \langle\hat{\eta}_b\rangle +  \langle\hat{\eta}_c\rangle = 1,
\end{equation}
where $\langle\hat{\eta}_a\rangle$, $\langle\hat{\eta}_b\rangle$, and $\langle\hat{\eta}_c\rangle$ are the probabilities to find the atom in the top, intermediate, and bottom levels, respectively. Therefore, applying Eqs. \eqref{94} and \eqref{95},  we arrive at
\begin{equation}\label{97}
 \langle\hat{\eta}_a\rangle = \frac{4\varepsilon^2}{\kappa^2 + 4\varepsilon^2}
\end{equation}
and
\begin{equation}\label{98}
 \langle\hat{\eta}_c\rangle = \frac{\kappa^2}{\kappa^2 + 4\varepsilon^2}.
\end{equation}
Finally, combination of Eqs. \eqref{91}, \eqref{94}, and  \eqref{98} leads to
\begin{equation}\label{99}
 \langle\hat{\sigma}_c\rangle = \frac{2\varepsilon\kappa }{\kappa^2 + 4\varepsilon^2}.
\end{equation}
\section{Quadrature Squeezing}
We next proceed to calculate the quadrature squeezing of the two-mode cavity light with the reservoir noise operators in normal order. To this end, applying the large-time approximation scheme to Eq. \eqref{86}, we get
\begin{equation}\label{100}
 \langle\hat{\sigma}_b(t)\rangle = 0.
\end{equation}
Then employing the complex conjugate of this result, one can put Eq. \eqref{85} in the form
\begin{equation}\label{101}
\frac{d}{dt} \langle\hat{\sigma}_a\rangle = -\frac{1}{2}\frac{\kappa^2\gamma_c}{\kappa^2 - 4\varepsilon^2}\langle\hat{\sigma}_a(t)\rangle.
\end{equation}
With the atom considered to be initially in the bottom level, the solution of Eq. \eqref{101} turns out to be
\begin{equation}\label{102}
\langle\hat{\sigma}_a(t)\rangle = 0.
\end{equation}
Moreover, with the atom considered to be initially in the bottom level, the solution of Eq. \eqref{86} is found to be
\begin{equation}\label{103}
\langle\hat{\sigma}_b(t)\rangle = 0.
\end{equation}
Now substituting the complex conjugate of the expectation value of Eq. \eqref{28} into Eq. \eqref{11}, we get
\begin{equation}\label{104}
 \frac{d}{dt}\langle\hat{a}(t)\rangle = -\bigg(\frac{\kappa^2 - 4\varepsilon^2}{2\kappa}\bigg)\langle\hat{a}(t)\rangle + g \bigg( \frac{2\varepsilon}{\kappa}\langle\hat{\sigma}^{\dagger}_b(t)\rangle - \langle\hat{\sigma}_a(t)\rangle\bigg)
\end{equation}
and on taking into account Eq. \eqref{100}, we have
\begin{equation}\label{105}
 \frac{d}{dt}\langle\hat{a}(t)\rangle = -\bigg(\frac{\kappa^2 - 4\varepsilon^2}{2\kappa}\bigg)\langle\hat{a}(t)\rangle - g\langle\hat{\sigma}_a(t)\rangle.
\end{equation}
The solution of this equation is a well-behaved function provided that 
\begin{equation}\label{106}
 \frac{\kappa^2 - 4\varepsilon^2}{2\kappa} > 0.
\end{equation}
It then follows that
\begin{equation}\label{107}
 \varepsilon < \frac{\kappa}{2}.
 \end{equation}
We realize that the the solution of Eq. \eqref{105} is expressible for $\varepsilon < \frac{\kappa}{2}$ as
\begin{eqnarray}\label{108}
 \langle\hat{a}(t)\rangle &=& \langle\hat{a}(0)\rangle exp\bigg(-\big(\frac{\kappa^2 - 4\varepsilon^2}{2\kappa}\big)t\bigg)\nonumber\\&& - g\int^{t}_{0} exp\bigg(-\big(\frac{\kappa^2 - 4\varepsilon^2}{2\kappa}\big)(t-t')\bigg)\langle\hat{\sigma}_a(t')\rangle dt'.
\end{eqnarray}
Hence with the assumption that the cavity light is initially in a vacuum state and in view of Eq. \eqref{102}, Eq. \eqref{108} turns out to be 
\begin{equation}\label{109}
\langle\hat{a}(t)\rangle = 0.
\end{equation}
Moreover, applying the large-time approximation scheme to Eq. \eqref{85}, we obtain
\begin{equation}\label{110}
\langle\hat{\sigma}_a(t)\rangle = \frac{2\varepsilon}{\kappa}\langle\hat{\sigma}^{\dagger}_b(t)\rangle.
\end{equation}
Furthermore, using the complex conjugate of the expectation value of Eq. \eqref{27} along with Eq. \eqref{110}, one can write Eq.\eqref{12} as
\begin{equation}\label{111}
\frac{d}{dt}\langle\hat{b}(t)\rangle = -\bigg(\frac{\kappa^2 - 4\varepsilon^2}{2\kappa}\bigg)\langle\hat{b}(t)\rangle- \frac{g}{\kappa^2}\bigg(\kappa^2 - 2\varepsilon^2\bigg)\langle\hat{\sigma}_b(t)\rangle.
\end{equation}
In view of Eq. \eqref{103} and the assumption that the cavity light is initially in a vacuum state, the solution of Eq. \eqref{111} turns out to be
\begin{equation}\label{112}
\langle\hat{b}(t)\rangle = 0.
\end{equation}
\indent
Now on adding Eqs. \eqref{105} and \eqref{111}, we have
\begin{equation}\label{113}
\frac{d}{dt}\langle\hat{c}(t)\rangle = -\bigg(\frac{\kappa^2 - 4\varepsilon^2}{2\kappa}\bigg)\langle\hat{c}(t)\rangle- g\langle\hat{\sigma}'(t)\rangle,
\end{equation}
where $\hat{c}(t)$ is the annihilation operator for the two-mode cavity light defined as 
\begin{equation}\label{114}
\hat{c}(t) = \hat{a}(t) + \hat{b}(t)
\end{equation}
and 
\begin{equation}\label{115}
\hat{\sigma}'(t) = \hat{\sigma}_a(t) + \frac{1}{\kappa^2}\bigg(\kappa^2 - 2\varepsilon^2\bigg)\langle\hat{\sigma}_b(t)\rangle
\end{equation}
On account of Eqs. \eqref{109} and \eqref{112}, Eq. \eqref{114} takes the form
\begin{equation}\label{116}
\langle\hat{c}(t)\rangle = 0.
\end{equation}
Now one can rewrite Eq. \eqref{113} as
\begin{equation}\label{117}
 \frac{d}{dt}\hat{c}(t) = -\frac{\beta}{2}\hat{c}(t) -g\hat{\sigma}'(t) + \hat{F}_c(t),
\end{equation}
where
\begin{equation}\label{118}
 \beta = \frac{\kappa^2 - 4\varepsilon^2}{\kappa}
\end{equation}
and $\hat{F}_c$ is the noise operator associated with the cavity mode operator $\hat{c}$ . Then we realize that the solution of Eq. \eqref{117} is given by
\begin{eqnarray}\label{119}
 \hat{c}(t)&=& \hat{c}(0)exp\bigg(-\frac{1}{2}\beta t\bigg) - g\int^{t}_{0}dt'exp\bigg(-\frac{1}{2}\beta(t - t')\bigg)\nonumber\\&& \times \bigg[\hat{\sigma}'(t') - \frac{1}{g}\hat{F}_c(t')\bigg]
\end{eqnarray}
and taking the adjoint of this equation, we have
\begin{eqnarray}\label{120}
 \hat{c}^{\dagger}(t)&=& \hat{c}^{\dagger}(0)exp\bigg(-\frac{1}{2}\beta t\bigg) - g\int^{t}_{0}dt'exp\bigg(-\frac{1}{2}\beta(t - t')\bigg)\nonumber\\&& \times \bigg[\hat{\sigma}'^{\dagger}(t') - \frac{1}{g}\hat{F}^{\dagger}_c(t')\bigg].
\end{eqnarray}
Upon multiplying Eqs.\eqref{119} and \eqref{120} by $\hat{F}_c(t)$ and $\hat{F}^{\dagger}_c(t)$ from the right side and taking the expectation values of the resulting equations, we get
\begin{eqnarray}\label{121}
 \langle\hat{c}(t)\hat{F}^{\dagger}_c(t)\rangle&=& \langle\hat{c}(0)\hat{F}^{\dagger}_c(t)\rangle exp\bigg(-\frac{1}{2}\beta t\bigg) - g\int^{t}_{0}dt'exp\bigg(-\frac{1}{2}\beta(t - t')\bigg)\nonumber\\&& \times \bigg[\langle\hat{\sigma}'(t')\hat{F}^{\dagger}_c(t)\rangle - \frac{1}{g}\langle\hat{F}_c(t')\hat{F}^{\dagger}_c(t)\rangle\bigg]
\end{eqnarray}
and
\begin{eqnarray}\label{122}
 \langle\hat{c}^{\dagger}(t)\hat{F}_c(t)\rangle &=& \langle\hat{c}^{\dagger}(0)\hat{F}_c(t)\rangle exp\bigg(-\frac{1}{2}\beta t\bigg) - g\int^{t}_{0}dt'exp\bigg(-\frac{1}{2}\beta(t - t')\bigg)\nonumber\\&& \times \bigg[\langle\hat{\sigma}'^{\dagger}(t')\hat{F}_c(t)\rangle - \frac{1}{g}\langle\hat{F}^{\dagger}_c(t')\hat{F}_c(t)\rangle\bigg].
\end{eqnarray}
On the basis of the fact that the noise operators $\hat{F}_c(t)$ at a certain time does not affect the cavity mode operator at earlier time, we can write
\begin{equation}\label{123}
 \langle\hat{c}^{\dagger}(0)\hat{F}_c(t)\rangle = \langle\hat{c}(0)\hat{F}^{\dagger}_c(t)\rangle = 0.
\end{equation}
With the help of Eqs. \eqref{43} and \eqref{123}, Eqs. \eqref{120} and \eqref{121} can be put in the form
\begin{eqnarray}\label{124}
 \langle\hat{c}(t)\hat{F}^{\dagger}_c(t)\rangle &=&  \int^{t}_{0}dt'exp\bigg(-\frac{1}{2}\beta(t - t')\bigg)\big\langle\hat{F}_c(t')\hat{F}^{\dagger}_c(t)\big\rangle
\end{eqnarray}
and 
\begin{eqnarray}\label{125}
 \langle\hat{c}^{\dagger}(t)\hat{F}_c(t)\rangle &=&  \int^{t}_{0}dt'exp\bigg(-\frac{1}{2}\beta(t - t')\bigg)\big\langle\hat{F}^{\dagger}_c(t')\hat{F}_c(t)\big\rangle.
\end{eqnarray}
\indent
The squeezing properties of the two-mode cavity light are described by two quadrature operators defined by
\begin{equation}\label{126}
 \hat{c}_+ = \hat{c}^{\dagger} + \hat{c}
\end{equation}
and
\begin{equation}\label{127}
 \hat{c}_- = i(\hat{c}^{\dagger} - \hat{c}).
\end{equation}
Then the variance of the quadrature operators is expressible as
 \begin{equation}\label{128}
  (\Delta c\pm)^2 = \pm\langle(\hat{c}^{\dagger} \pm \hat{c})^2\rangle \mp(\langle\hat{c}^{\dagger}\rangle \pm \langle\hat{c}\rangle)^2
 \end{equation}
and on account of Eq. \eqref{116}, we have
\begin{equation}\label{129}
  (\Delta c\pm)^2 = \langle\hat{c}^{\dagger}\hat{c}\rangle + \langle\hat{c}\hat{c}^{\dagger}\rangle \pm\big( \langle\hat{c}^{\dagger2}\rangle + \langle\hat{c}^2\rangle\big).
 \end{equation}
  This is the quadrature variance with the vacuum reservoir noise operators in arbitrary order. Applying Eq. \eqref{114} along with Eq. \eqref{68}, one readily establishes that
\begin{equation}\label{130}
 \langle\hat{c}^2\rangle = \big\langle\hat{a}\hat{b} + \hat{b}\hat{a}\big\rangle,
 \end{equation} 
so that on the basis of Eqs. \eqref{58} and \eqref{59} along with Eqs. \eqref{71} and \eqref{72}, this expression turns out to be
\begin{eqnarray}\label{131}
\langle\hat{c}^2\rangle &=& -\frac{2\varepsilon}{\kappa} - \frac{2\varepsilon}{\kappa}\bigg(\frac{4\varepsilon^2}{\kappa^2 - 4\varepsilon^2}\bigg) - \frac{\kappa\gamma_c}{(\kappa^2 - 4\varepsilon^2)^2}\bigg(3\varepsilon\kappa\langle\hat{\eta}_a\rangle + 4\varepsilon\kappa\langle\hat{\eta}_b\rangle + \varepsilon\kappa\langle\hat{\eta}_c\rangle \nonumber\\&&- [4\varepsilon^2\langle\hat{\sigma}_c + \hat{\sigma}^{\dagger}_c\rangle + ( \kappa^2 -4\varepsilon^2)\langle\hat{\sigma}_c\rangle]\bigg).
\end{eqnarray}\\
\indent
Now we proceed to calculate the quadrature variance by normally ordering the reservoir noise operators. Thus the normally ordered quadrature variance can be expressed as
\begin{equation}\label{132}
 \bigg(:(\Delta c_{\pm})^2:\bigg)_{F} = \langle:\hat{c}^{\dagger}\hat{c}:\rangle_F + \langle:\hat{c}\hat{c}^{\dagger}:\rangle_F \pm \langle :\hat{c}^2: + :\hat{c}^{\dagger2}:\rangle_F,
\end{equation}
where $(::)_F$ stands for normal ordering of the reservoir noise operators. Here we need to determine the various expectation values that appear in this equation. To this end, upon adding Eqs. \eqref{23} and \eqref{24}, one easily finds 
\begin{eqnarray}\label{133}
 \frac{d}{dt}\hat{c}(t) = -\frac{\kappa}{2}\hat{c}(t) -\varepsilon\hat{c}^{\dagger}(t)
 - g\hat{\sigma}(t) + \hat{F}_c(t),
\end{eqnarray}
with the adjoint of this equation given by
\begin{eqnarray}\label{134}
\frac{d}{dt}\hat{c}^{\dagger}(t) = -\frac{\kappa}{2}\hat{c}^{\dagger}(t) -\varepsilon\hat{c}(t)
 - g\hat{\sigma}^{\dagger}(t) + \hat{F}^{\dagger}_c(t),
\end{eqnarray}
where
\begin{eqnarray}\label{135}
 \hat{\sigma}(t) = \hat{\sigma}_a(t) + \hat{\sigma}_b(t)
\end{eqnarray}
and 
\begin{eqnarray}\label{136}
 \hat{F}_c(t) = \hat{F}_a(t) + \hat{F}_b(t).
\end{eqnarray}
Employing the relation
\begin{eqnarray}\label{137}
 \frac{d}{dt}\langle\hat{c}^{\dagger}(t)\hat{c}(t)\rangle = \langle\frac{d\hat{c}^{\dagger}(t)}{dt}\hat{c}(t)\rangle + \langle\hat{c}^{\dagger}(t)\frac{d\hat{c}(t)}{dt}\rangle 
\end{eqnarray}
together with Eqs. \eqref{133} and \eqref{134}, we arrive at
\begin{eqnarray}\label{138}
  \frac{d}{dt}\langle\hat{c}^{\dagger}(t)\hat{c}(t)\rangle &=& -\kappa\langle\hat{c}^{\dagger}(t)\hat{c}(t)\rangle - \varepsilon\bigg(\big\langle \hat{c}^{\dagger2}(t)\big\rangle + \big\langle\hat{c}^{2}(t)\big\rangle\bigg) - g\bigg(\big\langle \hat{c}^{\dagger}(t)\hat{\sigma}(t)\big\rangle\nonumber\\&& + \big\langle\hat{\sigma}^{\dagger}(t)\hat{c}(t)\big\rangle\bigg) + \big\langle \hat{c}^{\dagger}(t)\hat{F}_c(t)\big\rangle + \big\langle\hat{F}^{\dagger}_c(t)\hat{c}(t)\big\rangle.
\end{eqnarray}
This equation can be rewritten as
\begin{eqnarray}\label{139}
  \frac{d}{dt}\langle:\hat{c}^{\dagger}(t)\hat{c}(t):\rangle_F &=& -\kappa\langle:\hat{c}^{\dagger}(t)\hat{c}(t):\rangle_F - \varepsilon\bigg(\big\langle:\hat{c}^{\dagger2}(t):\big\rangle_F + \big\langle:\hat{c}^{2}(t):\big\rangle_F\bigg) \nonumber\\&&- g\bigg(\big\langle:\hat{c}^{\dagger}(t)\hat{\sigma}(t):\big\rangle_F + \big\langle:\hat{\sigma}^{\dagger}(t)\hat{c}(t):\big\rangle_F\bigg)\nonumber\\&& + \big\langle:\hat{c}^{\dagger}(t)\hat{F}_c(t):\big\rangle_F + \big\langle:\hat{F}^{\dagger}_c(t)\hat{c}(t):\big\rangle_F.
\end{eqnarray}
Furthermore, on applying the large-time approximation scheme to Eqs. \eqref{133} and \eqref{134}, we find the approximately valid relations
\begin{eqnarray}\label{140}
 \hat{c}(t) = -\frac{2\varepsilon}{\kappa}\hat{c}^{\dagger}(t)
 - \frac{2g}{\kappa}\hat{\sigma}(t) + \frac{2}{\kappa}\hat{F}_c(t)
\end{eqnarray}
and
\begin{eqnarray}\label{141}
\hat{c}^{\dagger}(t) = -\frac{2\varepsilon}{\kappa}\hat{c}(t)
 - \frac{2g}{\kappa}\hat{\sigma}^{\dagger}(t) + \frac{2}{\kappa}\hat{F}^{\dagger}_c(t).
\end{eqnarray}
Then one can easily establish that
\begin{equation}\label{142}
 \hat{c}(t) = \frac{4\kappa g}{\kappa^2 - 4\varepsilon^2}\bigg(\frac{\varepsilon\hat{\sigma}^{\dagger}(t)}{\kappa} - \frac{\hat{\sigma}(t)}{2} + \frac{\hat{F}_c(t)}{2g} - \frac{\varepsilon\hat{F}^{\dagger}_c(t)}{\kappa g}\bigg).
\end{equation}
In view of this expression and its adjoint, we have
\begin{eqnarray}\label{143}
\big\langle:\hat{c}^{\dagger}(t)\hat{\sigma}(t):\big\rangle_F + \big\langle:\hat{\sigma}^{\dagger}(t)\hat{c}(t):\big\rangle_F &=& \frac{4\kappa g}{\kappa^2 - 4\varepsilon^2}\bigg[\frac{\varepsilon\langle\hat{\sigma}(t)\hat{\sigma}(t)\rangle}{\kappa} - \frac{\langle\hat{\sigma}^{\dagger}(t)\hat{\sigma}(t)\rangle}{2}\nonumber\\&& + \frac{\langle\hat{F}^{\dagger}_c(t)\hat{\sigma}(t)\rangle}{2g} - \frac{\varepsilon\langle\hat{F}_c(t)\hat{\sigma}(t)\rangle}{\kappa g}\nonumber\\&& + \frac{\varepsilon\langle\hat{\sigma}^{\dagger}(t)\hat{\sigma}^{\dagger}(t)\rangle}{\kappa} - \frac{\langle\hat{\sigma}^{\dagger}(t)\hat{\sigma}(t)\rangle}{2} \nonumber\\&&+ \frac{\langle\hat{\sigma}^{\dagger}(t)\hat{F}_c(t)\rangle}{2g} - \frac{\varepsilon\langle\hat{\sigma}^{\dagger}(t)\hat{F}^{\dagger}_c(t)\rangle}{\kappa g}\bigg].
\end{eqnarray}
Thus on account of Eq. \eqref{43}, Eq. \eqref{143} takes the form
\begin{eqnarray}\label{144}
\big\langle:\hat{c}^{\dagger}(t)\hat{\sigma}(t):\big\rangle_F + \big\langle:\hat{\sigma}^{\dagger}(t)\hat{c}(t):\big\rangle_F &=& \frac{4\kappa g}{\kappa^2 - 4\varepsilon^2}\bigg[\frac{\varepsilon\langle\hat{\sigma}(t)\hat{\sigma}(t)\rangle}{\kappa} - \frac{\langle\hat{\sigma}^{\dagger}(t)\hat{\sigma}(t)\rangle}{2}\nonumber\\&& + \frac{\varepsilon\langle\hat{\sigma}^{\dagger}(t)\hat{\sigma}^{\dagger}(t)\rangle}{\kappa} - \frac{\langle\hat{\sigma}^{\dagger}(t)\hat{\sigma}(t)\rangle}{2}\bigg].
\end{eqnarray}
Making use of Eq. \eqref{135}, one can readily establish that
\begin{equation}\label{145}
 \frac{\varepsilon\langle\hat{\sigma}(t)\hat{\sigma}(t)\rangle}{\kappa} - \frac{\langle\hat{\sigma}^{\dagger}(t)\hat{\sigma}(t)\rangle}{2} = \frac{\varepsilon}{\kappa}\bigg\langle\hat{\sigma}_c(t)\bigg\rangle - \frac{1}{2}\bigg\langle \hat{\eta}_a(t) + \hat{\eta}_b(t)\bigg\rangle
\end{equation}
and
\begin{eqnarray}\label{146}
\frac{\varepsilon\langle\hat{\sigma}^{\dagger}(t)\hat{\sigma}^{\dagger}(t)\rangle}{\kappa} - \frac{\langle\hat{\sigma}^{\dagger}(t)\hat{\sigma}(t)\rangle}{2} = \frac{\varepsilon}{\kappa}\bigg\langle\hat{\sigma}^{\dagger}_c(t)\bigg\rangle - \frac{1}{2}\bigg\langle \hat{\eta}_a(t) + \hat{\eta}_b(t)\bigg\rangle.
\end{eqnarray}
Now on account of Eqs. \eqref{93} and \eqref{94}, Eqs. \eqref{145} and \eqref{146} turn out to be
\begin{equation}\label{147}
 \frac{\varepsilon\langle\hat{\sigma}(t)\hat{\sigma}(t)\rangle}{\kappa} - \frac{\langle\hat{\sigma}^{\dagger}(t)\hat{\sigma}(t)\rangle}{2} = 0
\end{equation}
and
\begin{eqnarray}\label{148}
\frac{\varepsilon\langle\hat{\sigma}^{\dagger}(t)\hat{\sigma}^{\dagger}(t)\rangle}{\kappa} - \frac{\langle\hat{\sigma}^{\dagger}(t)\hat{\sigma}(t)\rangle}{2} = 0.
\end{eqnarray}
Then in view of the results given by Eqs. \eqref{147} and \eqref{148}, Eq. \eqref{144} becomes
\begin{eqnarray}\label{149}
\big\langle:\hat{c}^{\dagger}(t)\hat{\sigma}(t):\big\rangle_F + \big\langle:\hat{\sigma}^{\dagger}(t)\hat{c}(t):\big\rangle_F = 0.
\end{eqnarray}
\indent
Upon rewriting Eq. \eqref{125}, we have 
\begin{eqnarray}\label{150}
 \langle:\hat{c}^{\dagger}(t)\hat{F}_c(t):\rangle_F &=&  \int^{t}_{0}dt'exp\bigg(-\frac{1}{2}\beta(t - t')\bigg)\big\langle:\hat{F}^{\dagger}_c(t')\hat{F}_c(t):\big\rangle_F.
\end{eqnarray}
Hence we see that
\begin{eqnarray}\label{151}
 \langle:\hat{c}^{\dagger}(t)\hat{F}_c(t):\rangle_F &=&  \int^{t}_{0}dt'exp\bigg(-\frac{1}{2}\beta(t - t')\bigg)\big\langle\hat{F}^{\dagger}_c(t)\hat{F}_c(t')\big\rangle.
\end{eqnarray}
We recall that for a vacuum reservoir
\begin{eqnarray}\label{152}
 \big\langle\hat{F}^{\dagger}_c(t)\hat{F}_c(t')\big\rangle_F = 0.
\end{eqnarray}
Thus Eq. \eqref{151} turns out to be 
\begin{eqnarray}\label{153}
 \langle:\hat{c}^{\dagger}(t)\hat{F}_c(t):\rangle &=&0.
\end{eqnarray}
Hence on account of Eqs. \eqref{149} and \eqref{153} along with its complex conjugate, Eq. \eqref{139} takes the form
\begin{eqnarray}\label{154}
  \frac{d}{dt}\langle:\hat{c}^{\dagger}(t)\hat{c}(t):\rangle_F &=& -\kappa\langle:\hat{c}^{\dagger}(t)\hat{c}(t):\rangle_F - \varepsilon\big\langle:\hat{c}^{\dagger2}(t): + :\hat{c}^{2}(t):\big\rangle.
\end{eqnarray}
Now the steady-state solution of the above equation is found to be 
\begin{eqnarray}\label{155}
  \langle:\hat{c}^{\dagger}\hat{c}:\rangle_F &=& -\frac{\varepsilon}{\kappa}\bigg\langle:\hat{c}^{\dagger2}: + :\hat{c}^{2}:\bigg\rangle_F.
\end{eqnarray}
The effect of the noise operators doesn't appear in this equation. This is because we have calculated the result given by Eq. \eqref{155} with the cavity mode coupled to a vacuum reservoir and with the noise operators in normal order.\\
\indent
Furthermore, upon applying the relation
\begin{eqnarray}\label{156}
 \frac{d}{dt}\langle\hat{c}(t)\hat{c}^{\dagger}(t)\rangle = \langle\hat{c}(t)\frac{d\hat{c}^{\dagger}(t)}{dt}\rangle + \langle\frac{d\hat{c}(t)}{dt}\hat{c}^{\dagger}(t)\rangle 
\end{eqnarray}
together with Eqs. \eqref{132} and \eqref{133}, we easily find
\begin{eqnarray}\label{157}
  \frac{d}{dt}\langle\hat{c}(t)\hat{c}^{\dagger}(t)\rangle &=& -\kappa\langle\hat{c}(t)\hat{c}^{\dagger}(t)\rangle - \varepsilon\bigg(\big\langle \hat{c}^{\dagger2}(t)\big\rangle + \big\langle\hat{c}^{2}(t)\big\rangle\bigg)\nonumber\\&& - g\bigg(\big\langle\hat{\sigma}(t)\hat{c}^{\dagger}(t)\big\rangle + \big\langle\hat{c}(t)\hat{\sigma}^{\dagger}(t)\big\rangle\bigg)\nonumber\\&& + \big\langle\hat{F}_c(t)\hat{c}^{\dagger}(t)\big\rangle + \big\langle\hat{c}(t)\hat{F}^{\dagger}_c(t)\big\rangle.
\end{eqnarray}
We can then write
\begin{eqnarray}\label{158}
  \frac{d}{dt}\langle:\hat{c}(t)\hat{c}^{\dagger}(t):\rangle_F &=& -\kappa\langle:\hat{c}(t)\hat{c}^{\dagger}(t):\rangle_F - \varepsilon\bigg(\big\langle:\hat{c}^{\dagger2}(t):\big\rangle_F + \big\langle:\hat{c}^{2}(t):\big\rangle_F\bigg)\nonumber\\&& - g\bigg(\big\langle:\hat{\sigma}(t)\hat{c}^{\dagger}(t):\big\rangle_F + \big\langle:\hat{c}(t)\hat{\sigma}^{\dagger}(t):\big\rangle_F\bigg)\nonumber\\&& + \big\langle :\hat{F}_c(t)\hat{c}^{\dagger}(t):\big\rangle_F + \big\langle:\hat{c}(t)\hat{F}^{\dagger}_c(t):\big\rangle_F.
\end{eqnarray}
Upon using Eq. \eqref{141} and its complex conjugate, we have
\begin{eqnarray}\label{159}
\big\langle:\hat{\sigma}(t)\hat{c}^{\dagger}(t):\big\rangle_F + \big\langle:\hat{c}(t)\hat{\sigma}^{\dagger}(t):\big\rangle_F &=& \frac{4\kappa g}{\kappa^2 - 4\varepsilon^2}\bigg[\frac{\varepsilon\langle\hat{\sigma}(t)\hat{\sigma}(t)\rangle}{\kappa} - \frac{\langle\hat{\sigma}(t)\hat{\sigma}^{\dagger}(t)\rangle}{2}\nonumber\\&& + \frac{\langle\hat{\sigma}(t)\hat{F}^{\dagger}_c(t)\rangle}{2g} - \frac{\varepsilon\langle\hat{\sigma}(t)\hat{F}_c(t)\rangle}{\kappa g}\nonumber\\&& + \frac{\varepsilon\langle\hat{\sigma}^{\dagger}(t)\hat{\sigma}^{\dagger}(t)\rangle}{\kappa} - \frac{\langle\hat{\sigma}(t)\hat{\sigma}^{\dagger}(t)\rangle}{2} \nonumber\\&&+ \frac{\langle\hat{F}_c(t)\hat{\sigma}^{\dagger}(t)\rangle}{2g} - \frac{\varepsilon\langle\hat{F}^{\dagger}_c(t)\hat{\sigma}^{\dagger}(t)\rangle}{\kappa g}\bigg].
\end{eqnarray}
In addition, in view of Eq. \eqref{43}, Eq. \eqref{159} takes the form
\begin{eqnarray}\label{160}
\big\langle:\hat{\sigma}(t)\hat{c}^{\dagger}(t):\big\rangle_F + \big\langle:\hat{c}(t)\hat{\sigma}^{\dagger}(t):\big\rangle_F &=& \frac{4\kappa g}{\kappa^2 - 4\varepsilon^2}\bigg[\frac{\varepsilon\langle\hat{\sigma}(t)\hat{\sigma}(t)\rangle}{\kappa} - \frac{\langle\hat{\sigma}(t)\hat{\sigma}^{\dagger}(t)\rangle}{2}\nonumber\\&& + \frac{\varepsilon\langle\hat{\sigma}^{\dagger}(t)\hat{\sigma}^{\dagger}(t)\rangle}{\kappa} - \frac{\langle\hat{\sigma}(t)\hat{\sigma}^{\dagger}(t)\rangle}{2}\bigg].
\end{eqnarray}
Moreover, applying Eq. \eqref{134}, one can readily establish that
\begin{equation}\label{161}
 \frac{\varepsilon\langle\hat{\sigma}(t)\hat{\sigma}(t)\rangle}{\kappa} - \frac{\langle\hat{\sigma}(t)\hat{\sigma}^{\dagger}(t)\rangle}{2} = \frac{\varepsilon}{\kappa}\bigg\langle\hat{\sigma}_c(t)\bigg\rangle - \frac{1}{2}\bigg\langle \hat{\eta}_c(t) + \hat{\eta}_b(t)\bigg\rangle
\end{equation}
and
\begin{eqnarray}\label{162}
\frac{\varepsilon\langle\hat{\sigma}^{\dagger}(t)\hat{\sigma}^{\dagger}(t)\rangle}{\kappa} - \frac{\langle\hat{\sigma}(t)\hat{\sigma}^{\dagger}(t)\rangle}{2} = \frac{\varepsilon}{\kappa}\bigg\langle\hat{\sigma}^{\dagger}_c(t)\bigg\rangle - \frac{1}{2}\bigg\langle \hat{\eta}_c(t) + \hat{\eta}_b(t)\bigg\rangle.
\end{eqnarray}
Then taking into account Eqs. \eqref{93} and \eqref{94}, Eqs. \eqref{161} and \eqref{162} can be written as
\begin{equation}\label{163}
 \frac{\varepsilon\langle\hat{\sigma}(t)\hat{\sigma}(t)\rangle}{\kappa} - \frac{\langle\hat{\sigma}(t)\hat{\sigma}^{\dagger}(t)\rangle}{2} = \frac{1}{2}\bigg\langle\hat{\eta}_a(t) - \hat{\eta}_c(t)\bigg\rangle
\end{equation}
and
\begin{eqnarray}\label{164}
\frac{\varepsilon\langle\hat{\sigma}^{\dagger}(t)\hat{\sigma}^{\dagger}(t)\rangle}{\kappa} - \frac{\langle\hat{\sigma}(t)\hat{\sigma}^{\dagger}(t)\rangle}{2} = \frac{1}{2}\bigg\langle\hat{\eta}_a(t) - \hat{\eta}_c(t)\bigg\rangle,
\end{eqnarray}
where we have used the fact that $\langle\hat{\sigma}_c(t)\rangle = \langle\hat{\sigma}^{\dagger}_c(t)\rangle$. Hence on employing Eqs. \eqref{163} and \eqref{164} in Eq. \eqref{160}, we get
\begin{eqnarray}\label{165}
\big\langle:\hat{\sigma}(t)\hat{c}^{\dagger}(t):\big\rangle_F + \big\langle:\hat{c}(t)\hat{\sigma}^{\dagger}(t):\big\rangle_F  = \frac{4g\kappa}{\kappa^2 - 4\varepsilon^2}\bigg\langle\hat{\eta}_a(t) - \hat{\eta}_c(t)\bigg\rangle.
\end{eqnarray}
Furthermore, Eq. \eqref{122} can be put in the from
\begin{eqnarray}\label{166}
 \langle:\hat{c}(t)\hat{F}^{\dagger}_c(t):\rangle_F &=&  \int^{t}_{0}dt'exp\bigg(-\frac{1}{2}\beta(t - t')\bigg)\big\langle:\hat{F}_c(t')\hat{F}^{\dagger}_c(t):\big\rangle_F.
\end{eqnarray}
Thus we notice that
\begin{eqnarray}\label{167}
 \langle:\hat{c}(t)\hat{F}^{\dagger}_c(t):\rangle_F &=&  \int^{t}_{0}dt'exp\bigg(-\frac{1}{2}\beta(t - t')\bigg)\big\langle\hat{F}^{\dagger}_c(t)\hat{F}_c(t')\big\rangle.
\end{eqnarray}
In view of Eq. \eqref{152}, Eq. \eqref{167} goes over into
\begin{eqnarray}\label{168}
 \langle:\hat{c}(t)\hat{F}^{\dagger}_c(t):\rangle_F = 0.
\end{eqnarray}

Then employing Eqs. \eqref{164} and \eqref{168} along with its complex conjugate , Eq. \eqref{158} can be written as
\begin{eqnarray}\label{169}
  \frac{d}{dt}\langle:\hat{c}(t)\hat{c}(t)^{\dagger}:\rangle_F &=& -\kappa\langle:\hat{c}(t)\hat{c}^{\dagger}(t):\rangle_F - \varepsilon\big\langle \hat{c}^{\dagger2}(t) + \hat{c}^{2}(t)\big\rangle \nonumber\\&&- \frac{\kappa^2\gamma_c}{\kappa^2 - 4\varepsilon^2}\bigg\langle\hat{\eta}_a(t) - \hat{\eta}_c(t)\bigg\rangle.
\end{eqnarray}
The steady-state solution of the above equation is found to be 
\begin{eqnarray}\label{170}
  \langle:\hat{c}\hat{c}^{\dagger}:\rangle_F &=& -\frac{\varepsilon}{\kappa}\bigg\langle:\hat{c}^{\dagger2}: + :\hat{c}^{2}:\bigg\rangle_F - \frac{\kappa\gamma_c}{\kappa^2 - 4\varepsilon^2}\bigg\langle\hat{\eta}_a - \hat{\eta}_c\bigg\rangle.
\end{eqnarray}
The effect of the noise operators doesn't appear in this equation. This is because we have calculated the result given by Eq. \eqref{170} by normally ordering the noise operators, with the cavity mode coupled to a vacuum reservoir. Now upon introducing Eqs. \eqref{155} and \eqref{170} into Eq. \eqref{132}, we find
\begin{equation}\label{171}
 \bigg(:(\Delta c_+)^2:\bigg)_{F}  =  \frac{\kappa\gamma_c}{\kappa^2 - 4\varepsilon^2}\bigg\langle\hat{\eta}_c - \hat{\eta}_a\bigg\rangle + \bigg\langle:\hat{c}^{\dagger2}: + :\hat{c}^{2}:\bigg\rangle\bigg(-\frac{2\varepsilon}{\kappa} + 1\bigg)
\end{equation} 
and 
\begin{equation}\label{172}
 \bigg(:(\Delta c_-)^2:\bigg)_{F}  =  \frac{\kappa\gamma_c}{\kappa^2 - 4\varepsilon^2}\bigg\langle\hat{\eta}_c - \hat{\eta}_a\bigg\rangle - \bigg\langle:\hat{c}^{\dagger2}: + :\hat{c}^{2}:\bigg\rangle_F\bigg(\frac{2\varepsilon}{\kappa} + 1\bigg).
\end{equation}
In addition, upon substituting \eqref{131} and its complex conjugate along with Eqs. \eqref{94}, and \eqref{97}-\eqref{99} into these equations, we readily arrive at
\begin{equation}\label{173}
 \bigg(:(\Delta c_+)^2:\bigg)_{F}  =  \frac{\kappa\gamma_c}{\kappa^2 + 4\varepsilon^2}\bigg(\frac{\kappa + 4\varepsilon}{\kappa + 2\varepsilon}\bigg) - \frac{4\varepsilon}{\kappa + 2\varepsilon}
\end{equation} 
and 
\begin{equation}\label{174}
 \bigg(:(\Delta c_-)^2:\bigg)_{F}  =   \frac{\kappa\gamma_c}{\kappa^2 + 4\varepsilon^2}\bigg(\frac{\kappa - 4\varepsilon}{\kappa - 2\varepsilon}\bigg) + \frac{4\varepsilon}{\kappa - 2\varepsilon}.
\end{equation}
These results represent the plus and minus quadrature variance for the two-mode cavity light obtained by normally ordering the noise operators. The first term in Eq. \eqref{173} or \eqref{174} represents the effect of the interaction of the subharmonic light modes with the three-level atom. Whereas the second term is due to the subharmonic generation. Moreover, we note that for $\varepsilon = 0$, Eqs.  reduce to
 \begin{equation}\label{175}
  \bigg(:(\Delta c_+)^2_{vac}:\bigg)_{F} = \bigg(:(\Delta c_-)^2_{vac}:\bigg)_{F} = \frac{\gamma_c}{\kappa}.
 \end{equation}
 This indeed represents the quadrature variance for a two-mode cavity vacuum state with the noise operators in normal order. From the Eqs. \eqref{173} and \eqref{174} we realize that the squeezing of the two-mode cavity light occurs in the plus quadrature variance. So, it appears to be interesting to find out the value of $\gamma_c$ for which the variance of plus quadrature is zero. Thus
 \begin{equation}\label{176}
  \frac{\kappa\gamma_c}{\kappa^2 + 4\varepsilon^2}\bigg(\frac{\kappa + 4\varepsilon}{\kappa + 2\varepsilon}\bigg) - \frac{4\varepsilon}{\kappa + 2\varepsilon} = 0
\end{equation} 
 Then for $\varepsilon \approx 0.4$, we easily get
 \begin{equation}\label{177}
  \gamma_c = \frac{16}{15} \approx 1.07
 \end{equation}
 \begin{figure*}
\centering
\begin{center}
\includegraphics[width=12cm, height = 9cm]{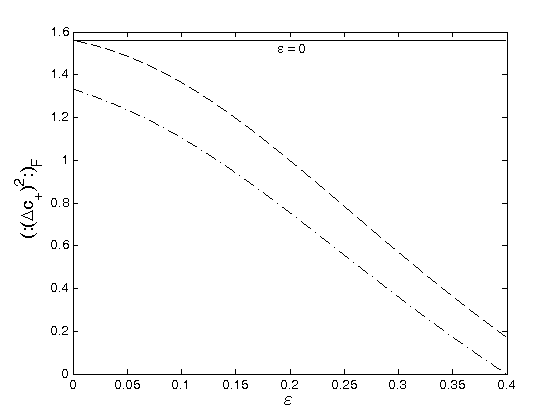}
\caption{\footnotesize{ Plots of the plus quadrature variance [Eqs. \eqref{173} and \eqref{175}] versus $\varepsilon$ for $\kappa = 0.8$, $\gamma_c = \frac{16}{15} \approx 1.07$ (dotted line), and $\gamma_c = 1.25$ (dashed line).}}
\end{center}
\end{figure*}
\begin{figure*}
\centering
\begin{center}
\includegraphics[width=12cm, height = 9cm]{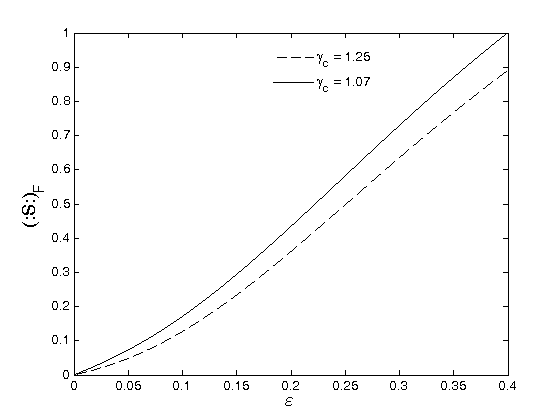}
\caption{\footnotesize{Plots of the quadrature squeezing [Eq. \eqref{180}] versus $\varepsilon$ for $\kappa = 0.8$, $\gamma_c = \frac{16}{15}$, and $\gamma_c = 1.25$.}}
\end{center}
\end{figure*}
From the plots in Fig.3, we see that the quadrature variance is below the vacuum state-level (at $\varepsilon = 0$) and decreases until it eventually approaches to zero (at $\gamma_c = \frac{16}{15} \approx 1.07$).\\ 
\indent
Now we calculate the quadrature squeezing of the two-mode cavity light with the reservoir noise operators in normal order. We define the quadrature squeezing by
\begin{equation}\label{178}
 \bigg(:S:\bigg)_F = \frac{\bigg(:(\Delta c_+)^2_{vac}:\bigg)_F - \bigg(:(\Delta c_+)^2:\bigg)_F}{\bigg(:(\Delta c_+)^2_{vac}:\bigg)_F}.
\end{equation}
It then follows that 
\begin{equation}\label{179}
 \bigg(:S:\bigg)_F = 1- \frac{\bigg(:(\Delta c_+)^2:\bigg)_F}{\bigg(:(\Delta c_+)^2_{vac}:\bigg)_F}.
\end{equation}
Then on applying Eqs. \eqref{175} and \eqref{177} in Eq. \eqref{179}, we readily get
 \begin{eqnarray}\label{180}
\bigg(:S:\bigg)_F &=& 1- \frac{\kappa}{\gamma_c}\bigg[ \frac{\kappa\gamma_c}{\kappa^2 + 4\varepsilon^2}\bigg(\frac{\kappa + 4\varepsilon}{\kappa + 2\varepsilon}\bigg) - \frac{4\varepsilon}{\kappa + 2\varepsilon}\bigg].
\end{eqnarray}
This represents the quadrature squeezing for the two-mode cavity light when the noise operators are in normal order. From the plots in Fig.3, we notice that the maximum quadrature squeezing is $100\%$ for $\gamma_c = \frac{16}{15} \approx 1.07$ and is $88.3\%$ for $\gamma_c = 1.25$ at values of $\varepsilon$ close to 0.4.
\section{Conclusion}
\noindent 
We have analyzed the interaction of subharmonic light modes, emerging from a nonlinear crystal driven by coherent light, with a three-level atom in a closed cavity coupled to a vacuum reservoir via a single port mirror. Applying the steady-state solutions of the equations of evolution for the cavity mode and atomic operators, we have calculated the quadrature squeezing by putting the vacuum reservoir noise operators in normal order. We found that the maximum quadrature squeezing is $100\%$ for $\gamma_c = \frac{16}{15} \approx 1.07$ and at values of $\varepsilon$ close to 0.4. It is not hard to realize that the increase in the quadrature squeezing is due to the normal ordering of the vacuum reservoir noise operators.
\section*{Affiliation}
Department of Physics, Addis Ababa University, P. O. Box 1176, Addis Ababa, Ethiopia\\ and Kotebe Metropolitan University.
\section*{Corresponding author}
Correspondence to : Merid Tufa
\section*{Conflict of interest and ethics declaration}
The author has no conflict of interest to declare that are relevant to the content of this article. Also no funding was received for conducting this study.

\end{document}